\documentclass[prd,twocolumn,floatfix,showpacs,preprintnumbers,nofootinbib]{revtex4}
\usepackage[utf8x]{inputenc}
\usepackage{mathrsfs}
\usepackage{amssymb}
\usepackage{amsmath}
\usepackage{graphicx}

\usepackage[caption = false]{subfig}
\usepackage[normalem]{ulem}
\usepackage{xcolor}
\definecolor{lcolor}{rgb}{0.5,0,0}
\definecolor{citcolor}{rgb}{0,0.3,0.0}

\usepackage[breaklinks,colorlinks,urlcolor=blue,citecolor=citcolor,linkcolor=lcolor]{hyperref}
\usepackage{mciteplus}

\newcommand{\rt}{{\mathbf{r}}}

\newcommand{\bt}{{\mathbf{b}}}
\newcommand{\bti}{{\mathbf{b}_{i}}}
\newcommand{\bqti}{{\mathbf{b}_{q,i}}}

\newcommand{\Deltat}{{\boldsymbol{\Delta}}}

\newcommand{\nc}{{N_\mathrm{c}}}

\newcommand{\gev}{\ \textrm{GeV}}
\newcommand{\fm}{\ \textrm{fm}}

\newcommand{\as}{\alpha_{\mathrm{s}}}

\newcommand{\eq}{Eq.~}

\newcommand{\xpom}{{x_\mathbb{P}}}

\newcommand{\der}{\mathrm{d}}

\newcommand{\A}{{\mathcal{A}}}

\begin{document}

\author{Heikki M\"antysaari}
\affiliation{
Physics Department, Brookhaven National Laboratory, Upton, NY 11973, USA
}

\author{Bj\"orn Schenke}
\affiliation{
Physics Department, Brookhaven National Laboratory, Upton, NY 11973, USA
}

\title{
Probing subnucleon scale fluctuations in ultraperipheral heavy ion collisions
}

\preprint{}

\begin{abstract}
We show that introducing subnucleon scale fluctuations constrained by HERA diffractive $J/\Psi$ production data significantly affects the incoherent diffractive $J/\Psi$ production cross section in ultraperipheral heavy ion collisions. We find that the inclusion of the additional fluctuations increases the ratio of the incoherent to the coherent cross section approximately by a factor of $2$, and modifies the transverse momentum spectra of the produced $J/\Psi$ at momenta larger than the scale that corresponds to the distance scale of the subnucleonic fluctuations.
We present predictions for  $J/\Psi$ production in ultraperipheral heavy ion collisions at $\sqrt{s_{NN}}=5.02\,{\rm TeV}$ at the LHC and $200\,{\rm GeV}$ at RHIC. 
\end{abstract}

\maketitle

\section{Introduction}

Deep inelastic scattering (DIS) is a clean process to study the internal structure of hadrons via interaction with (virtual) photons. The most precise data to date on the partonic structure of the proton comes from the DIS experiments performed at the HERA e$^\pm$ + p collider~\cite{Aaron:2009aa,Abramowicz:2015mha}. These measurements have shown that the gluonic density of the proton grows rapidly when the momentum fraction $x$ of the gluons decreases. 
Nonlinear QCD phenomena must limit this growth at very small $x$ in order to avoid violating unitarity.
These nonlinearities are most conveniently described within the Color Glass Condensate (CGC) effective theory of QCD~\cite{Iancu:2003xm,Gelis:2010nm}, which defines the framework we use in this work.

The gluon density scales as $A^{1/3}$, which amplifies nonlinear effects in heavy nuclei with large mass number $A$. However, currently there is no nuclear DIS data at small $x$ available. The proposed Electron Ion Collider (EIC)~ \cite{Boer:2011fh,Accardi:2012qut} and Large Hadron Electron Collider (LHeC)~\cite{AbelleiraFernandez:2012cc} are designed to explore this region of large gluon densities. Before these machines are realized, one possibility to study deep inelastic scattering on nuclei is given by ultraperipheral heavy ion collisions, where a large impact parameter suppresses the relatively short range strong interactions. Instead, scattering processes involve a photon, emitted by one of the electrically charged nuclei, scattering off the other nucleus. First results from the LHC on diffractive vector meson production in ultraperipheral collisions have demonstrated the sensitivity of this process on nuclear effects at small $x$~\cite{Abbas:2013oua,Abelev:2012ba,CMS:2014ies}. Recently, ultraperipheral proton-nucleus collisions, where the large electric charge of the nucleus causes the photon-proton scattering to dominate,  have also been used to study deep inelastic scattering off a proton at energies much higher than were available at HERA~\cite{TheALICE:2014dwa,CMS:2016nct}.

Diffractive scattering, where a system of particles or just a single particle is produced without exchanging a net color charge with the target, is a powerful process to study the small-$x$ structure of hadrons. At leading order in perturbative QCD, the diffractive cross section is proportional to the \emph{square} of the gluon density, making it very sensitive to small-$x$ gluons. In addition, exclusive vector meson production is sensitive to the geometric structure of the hadron. In particular, in coherent diffraction (where the target hadron remains in the same quantum state) an average density profile is probed. On the other hand, when the target breaks up, one is sensitive to the event-by-event fluctuations of the gluon fields in the target~\cite{Miettinen:1978jb,Frankfurt:1993qi,Frankfurt:2008vi,Kowalski:2008sa,Caldwell:2009ke,Lappi:2010dd}.  In recent phenomenological applications it has been demonstrated that one can indeed constrain the shape and the shape fluctuations of the proton (not included in previous literature) and nuclei at small $x$ by studying exclusive vector meson production~\cite{Toll:2012mb,Mantysaari:2016ykx,Cepila:2016uku} (see also~\cite{Alvioli:2014eda,Albacete:2016pmp,Welsh:2016siu,Mitchell:2016jio}). 
For a more formal introduction of diffractive scattering, the reader is referred to Ref.~\cite{Good:1960ba}.

We calculate coherent and incoherent diffractive $J/\Psi$ production in ultraperipheral heavy ion collisions in the dipole picture. 
The main focus of this work is to study how the diffractive cross sections are affected by inclusion of the subnucleon scale fluctuations that have been constrained using diffractive $J/\Psi$ production at HERA in Refs.~\cite{Mantysaari:2016ykx,Mantysaari:2016jaz}.
Apart from the inclusion of these fluctuations, we improve over previous state-of-the art CGC work~\cite{Lappi:2013am} (see also~\cite{Kovchegov:1999ji,kuroda:2005by,Goncalves:2005yr,Lappi:2014foa}) by using a Monte Carlo method to explicitly calculate target averages.  This allows us to use the
impact parameter dependent saturation model (IPsat) dipole amplitude~\cite{Kowalski:2003hm} without factorization of the impact parameter dependence (an approximation which was necessary to derive the incoherent cross section in Ref.~\cite{Lappi:2010dd}). We also use an updated IPsat parametrization fitted to the combined HERA data~\cite{Rezaeian:2012ji}.

\section{Dipole scattering}
\label{sec:dipole}

The basic ingredient in the dipole framework is the forward elastic dipole-target scattering amplitude $N(\rt,\bt,\xpom)$, where $\rt$ is the two dimensional vector that connects the quark and antiquark of the dipole in the transverse plane, $\bt$ is the impact parameter and $\xpom$ is the usual Bjorken variable of DIS in a diffractive event. In this work we use the IPsat model~\cite{Kowalski:2003hm}, which employs an eikonalized DGLAP evolved~\cite{Bartels:2002cj} gluon distribution $xg$ and includes saturation effects. The dipole-proton scattering amplitude is written as
\begin{equation}
\label{eq:ipsat_n}
 N(\rt,\bt,x) = 1 - \exp\left(- \frac{\pi^2 }{2\nc} \as(\mu^2) xg(x, \mu^2) \rt^2   T_p(\bt)\right) ,
\end{equation}
with the thickness function
\begin{equation}
	\label{eq:ipsat_tp}
	    T_p(\bt)=\frac{1}{2\pi B_p} e^{-\bt^2/(2B_p)}.
\end{equation}
Here both the coupling $\alpha_s$ and the gluon distribution are evaluated at the scale $\mu^2=\mu_0^2 + 4/\rt^2$. The proton width $B_p$, initial scale $\mu_0^2$ and the initial condition for the DGLAP evolution of the gluon distribution $x g$ are parameters of the model. Their values are determined in Ref.~\cite{Rezaeian:2012ji} by performing fits to HERA DIS data. 

To include proton structure fluctuations we follow Refs.~\cite{Mantysaari:2016ykx,Mantysaari:2016jaz} and assume that the gluonic density of the proton in the transverse plane is distributed around three constituent quarks (hot spots)\footnote{As shown in Ref.~\cite{Mantysaari:2016jaz}, it is also possible to describe the HERA data using a different number of hot spots, or tubes instead of round hot spots. Consequently, we do not expect our results to depend on the exact choice of model for the proton geometry, as long as it has the correct amount of fluctuations constrained by the HERA data.}. These hot spots are assumed to be Gaussian. In practice we perform the replacement of the impact parameter profile \eqref{eq:ipsat_tp}
\begin{equation}
\label{eq:fluctuations_replacement}
	T_p(\bt) \to \sum_{i=1}^3 T_q(\bt-\bqti) , \text{ with} \,\,\, T_q(\bt) = \frac{1}{2\pi B_q} e^{-\bt^2/(2B_q)}.
\end{equation}
where $\bqti$ are the locations of the hot spots. They are sampled from a two dimensional Gaussian distribution whose width is $B_{qc}$.  The free parameters $B_q$ and $B_{qc}$ are obtained in Ref.~\cite{Mantysaari:2016jaz} by comparing with the HERA coherent and incoherent diffractive $J/\Psi$ production data at the photon-proton center of mass energy $W \sim 75 \gev$, corresponding to $\xpom \sim 10^{-3}$.   The proton fluctuation parameters obtained are $B_{qc}=3.3\gev^{-2}$, $B_q=0.7\gev^{-2}$ and are close to the values obtained in a similar analysis in Ref.~\cite{Cepila:2016uku}.

An additional source of fluctuations we include here comes from fluctuations of the overall normalization of the saturation scale, which we refer to in short as saturation scale fluctuations. Following again Ref.~\cite{Mantysaari:2016jaz}, we allow the saturation scale of each of the constituent quarks to fluctuate independently according to a log-normal distribution. The width of that distribution is obtained in Refs.~\cite{McLerran:2015qxa,Bzdak:2015eii} by comparing to the $p+p$ multiplicity fluctuation data. The saturation scale fluctuations were shown in Ref.~\cite{Mantysaari:2016jaz} to be necessary to describe the incoherent diffractive cross section measured by HERA at small $|t|$. For a more detailed description of the implementation see Ref.~\cite{Mantysaari:2016jaz}.

The replacement \eqref{eq:fluctuations_replacement} changes the impact parameter dependence of the average dipole amplitude \eqref{eq:ipsat_n} even though we require $\langle \sum_i T_q(\bt-\bti)\rangle = T_p(\bt)$. 
Here, the average is taken over different nucleon configurations.
This is because the thickness function appears in the exponential. 
As a result, the description of the proton structure function data would not be as good as with the original fit~\cite{Rezaeian:2012ji}. Also, as shown in Ref.~\cite{Mantysaari:2016jaz}, this parametrization tends to slightly underestimate the coherent $\gamma p$ cross section measured at HERA. In principle one should perform a new fit to the HERA structure function data with the modified IPsat model parametrization. As the purpose of this work is to study the effect of the subnucleon scale fluctuations on ultraperipheral heavy ion collisions this fitting is left for future work.\footnote{There is also a large model uncertainty from the modeling of the vector meson wave function that will affect the overall normalization of the diffractive cross sections. We will return to this issue below and show that the ratio of incoherent to coherent cross sections is largely independent of the wave function.}

The dipole-nucleus amplitude $N_A$ is obtained by using  an independent scattering approximation, similar to Refs.~\cite{Kowalski:2003hm,Lappi:2010dd}
\begin{equation}
	N_A(\rt,\bt,x) = 1 - \prod_{i=1}^A \big[1 - N(\rt, \bt-\bti,x)\big],
\end{equation}
where $\bti$ are the transverse positions of the nucleons in the nucleus. The interpretation here is that $1-N$ is the probability not to scatter off an individual nucleon, thus $\prod_i (1-N_i)$ is the probability not to scatter off the entire nucleus.

\section{Diffractive deep inelastic scattering}
\label{sec:diffraction}

The scattering amplitude for diffractive vector meson production in $\gamma^*$-nucleus scattering can be written as~\cite{Kowalski:2006hc}
\begin{multline}
\label{eq:diff_amp}
 \A^{\gamma^* A \to V A}_{T,L}(\xpom,Q^2, \boldsymbol{\Delta}) = 2i\int \der^2 \rt \int \der^2 \bt \int \frac{\der z}{4\pi}  \\ 
 \times (\Psi^*\Psi_V)_{T,L}(Q^2, \rt,z) \\
 \times e^{-i[\bt - (1-z)\rt]\cdot \boldsymbol{\Delta}}  N_A(\rt,\bt,\xpom).
\end{multline}
Here the momentum transfer is $\Deltat=(P'-P)_\perp \approx \sqrt{-t}$, where $P$ and $P'$ are the momenta of the incoming and outgoing nucleus, respectively. The subscripts $T$ and $L$ refer to the transverse and longitudinal polarization of the virtual photon with virtuality $Q^2$. In ultraperipheral events, the photons are approximately real ($Q^2=0$) and only the transverse component contributes.

The scattering amplitude \eqref{eq:diff_amp} can be interpreted as follows: first, the incoming virtual photon fluctuates into a quark-antiquark dipole with transverse separation $|\rt|$, the quark carrying the momentum fraction $z$. This splitting is described by the  virtual photon wave function $\Psi$ (see e.g.~\cite{Kovchegov:2012mbw}). As discussed previously, the elastic scattering amplitude for the dipole to scatter off the nucleus is $N_A(\rt,\bt,\xpom)$. Finally, the vector meson $V$ is formed, and the $q\bar q \to V$ formation is described by the vector meson wave function $\Psi_V$. In this work we use both the Boosted Gaussian and Gaus-LC wave function parametrizations from Ref.~\cite{Kowalski:2006hc} in order to estimate the model uncertainty related to the formation of the $J/\Psi$
(see also Ref.~\cite{Chen:2016dlk} for a recent more rigorous calculation of the vector meson wave functions). 

Two phenomenological corrections to the diffractive cross sections are included. First,
equation \eqref{eq:diff_amp} 
is derived assuming that the dipole amplitude is completely real, which makes the diffractive scattering amplitude purely imaginary (in case of a rotationally symmetric dipole amplitude).
A correction for the presence of the real part is necessary. Secondly, the skewedness correction that takes into account the fact that in two-gluon exchange processes the gluons in the target are probed at different values of Bjorken-$x$ is also included. These corrections are discussed in more detail in Appendix~\ref{appendix:corrections}.

The coherent diffractive cross section is obtained by averaging the diffractive scattering amplitude over the target configurations and taking the square \cite{Miettinen:1978jb,Kowalski:2006hc}:
\begin{equation}
\label{eq:coherent}
\frac{\der \sigma^{\gamma^* A \to V A}}{\der t} = \frac{1}{16\pi} \left| \left\langle \A^{\gamma^* A \to V A}(\xpom,Q^2,\boldsymbol{\Delta}) \right\rangle \right|^2.
\end{equation}
Here the brackets $\langle \rangle$ refer to averages over different configurations of the target. The incoherent cross section is obtained by subtracting the coherent cross section from the total diffractive cross section. It takes the form of a variance of the diffractive scattering amplitude~\cite{Miettinen:1978jb} (see also Refs.~\cite{Frankfurt:1993qi,Frankfurt:2008vi,Caldwell:2009ke,Lappi:2010dd}):
\begin{align}\label{eq:incoherent}
\frac{\der \sigma^{\gamma^* A \to V A^*}}{\der t} = \frac{1}{16\pi} &\left( \left\langle \left| \A^{\gamma^* A \to V A}(\xpom,Q^2,\boldsymbol{\Delta})  \right|^2 \right\rangle \right. \notag\\ 
& ~~~ - \left. \left| \left\langle \A^{\gamma^* A \to V A}(\xpom,Q^2,\boldsymbol{\Delta}) \right\rangle \right|^2 \right)\,.
\end{align} 

The cross sections are related to Fourier transforms of the dipole-nucleus amplitude from  coordinate space to momentum space, and the transverse momentum transfer $\Deltat$ is the Fourier conjugate to $\bt - (1-z)\rt$. Here the impact parameter $\bt$ points to the center of the dipole from the center of the nucleus, and the factor $(1-z)\rt$ appears due to the use of non-forward wave functions~ \cite{Bartels:2003yj,Kowalski:2006hc}. 
The dependence on $\bt$ shows that
diffractive vector meson production is sensitive to the impact parameter profile,
 in contrast to calculations of proton structure functions where the impact parameter integral merely affects the overall normalization.  This makes diffractive scattering a sensitive probe of the internal geometric structure of hadrons and nuclei. In particular, larger momentum transfers probe  smaller distance scales, which we will show explicitly later. Because the incoherent cross section \eqref{eq:incoherent} has the form of a variance, it is sensitive to the amount of fluctuations in coordinate space.

In Ref.~\cite{Lappi:2010dd} the incoherent cross section was calculated analytically assuming that the impact parameter dependence of the dipole amplitude factorizes, and is explicitly proportional to $e^{-\bt^2/(2B_p)}$. In that case, the dipole amplitude does not saturate to unity at large dipoles or at large densities. As we calculate the target averages explicitly using a Monte Carlo method (similar to SARTRE~\cite{Toll:2012mb}), we do not need to rely on this approximation. We note that for the $J/\Psi$ production at the LHC, usage of the factorized approximation for the dipole amplitude reduces the coherent cross section by approximately $15\%$ when no subnucleonic fluctuations are included.

 The averages over target configurations are calculated by sampling hundreds configurations. This involves sampling nucleon positions from a Woods-Saxon distribution and the subnucleonic structure for each of the nucleons as described above.

The structure of the target is probed at the scale
\begin{equation}
\xpom = \frac{Q^2+M_V^2-t}{Q^2+W^2-m_N^2},
\end{equation}
which can be interpreted as the longitudinal momentum fraction of the nucleon carried by the color-neutral ``pomeron''. Here $W$ is the center-of-mass energy in the photon-nucleon scattering and $m_N$ and $M_V$ are  the nucleon and the vector meson masses, respectively.

\section{Exclusive vector meson production in ultraperipheral collisions}

Following Ref.~\cite{Bertulani:2005ru}, we write the vector meson production cross section in ultraperipheral heavy ion collisions as a convolution of photon flux generated by one of the nuclei $n^{A_i}$ and the photon-nucleus cross section $\sigma^{\gamma A_i}$:
\begin{equation}
\label{eq:upc}
\frac{\der \sigma^{AA \to J/\Psi AA'}}{\der t} = n^{A_2}(\omega_2 ) \sigma^{\gamma A_1}(y) + n^{A_1}(\omega_1) \sigma^{\gamma A_2}(-y) .
\end{equation}
Here $y$ is the rapidity of the vector meson, and the total photon flux $n^{A_i}(\omega)$ is obtained by integrating the photon flux of the nucleus over the impact parameter in the region $|\bt| > 2R_A$. For an explicit expression, the reader is referred to Ref.~\cite{Bertulani:2005ru}. The photon flux can be calculated by noticing that the photon energies are $\omega_1=(M_V/2)e^{y}$ and $\omega_2=(M_V/2)e^{-y}$. The center of mass energy squared of the photon-nucleon system is $W^2 = \sqrt{s_{NN}} M_V e^{\pm y}$, and the Bjorken-$x$ probed in the process becomes $\xpom \approx (M_V/\sqrt{s_{NN}}) e^{\mp y}$.

Note that there are two different contributions to the vector meson production at $y \neq 0$. Either a large-$x$ photon scatters off a small-$x$ gluon, or vice versa. This limits the applicability of the framework at forward and backward rapidities, where a significant contribution to the cross section comes from a process where the center-of-mass energy of the photon-nucleon scattering is small. Thus we shall only calculate the cross section in the region where $\xpom \lesssim 0.015$. At $\sqrt{s_{NN}}=2.76$ TeV this corresponds to $|y|\lesssim 2.5$, at $\sqrt{s_{NN}}=5.02$ TeV to $|y|\lesssim 3$ in the case of $J/\Psi$ production.

 \section{Results}

\begin{figure}[tb]
\centering
		\includegraphics[width=0.5\textwidth]{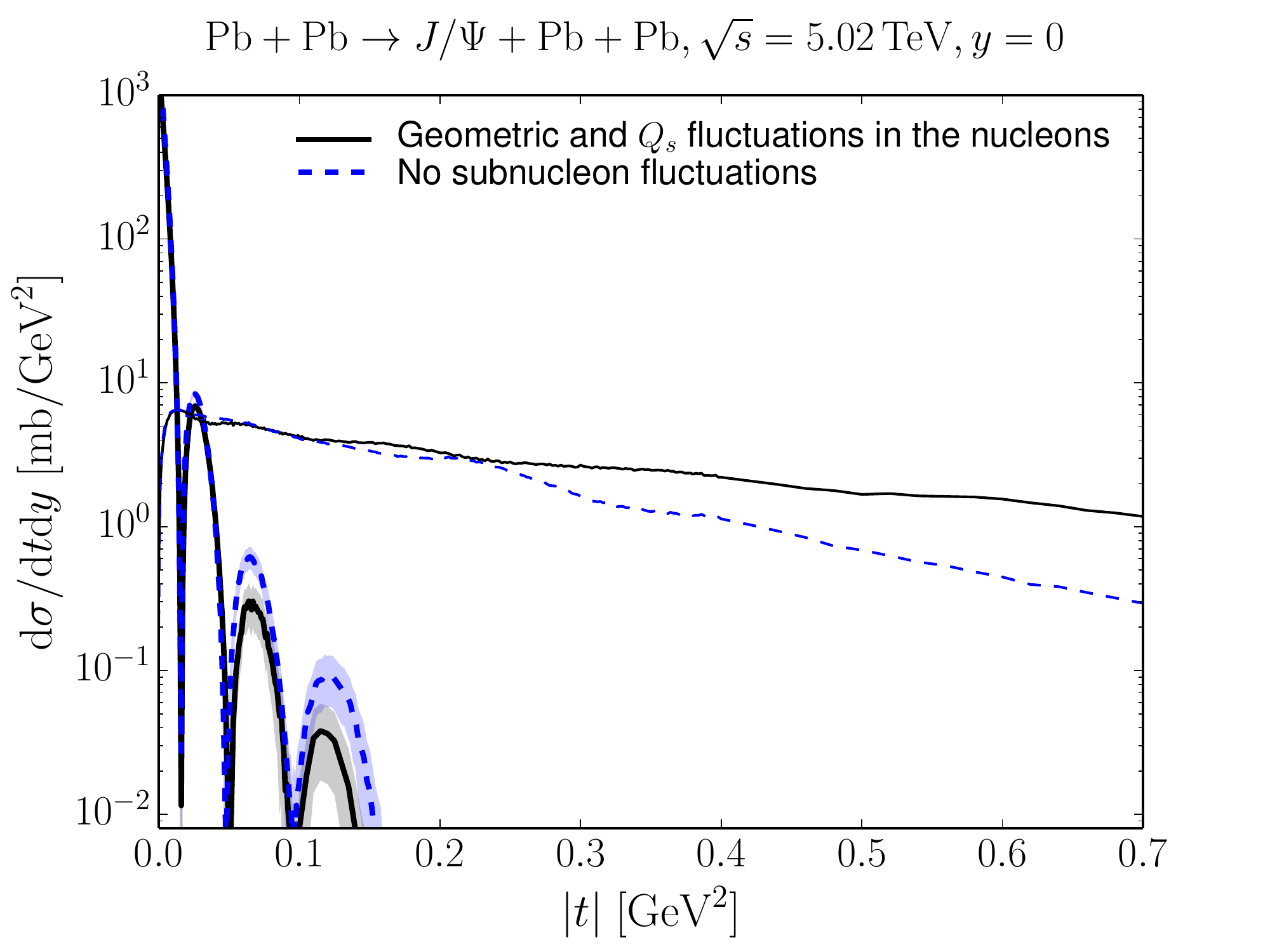} 
				\caption{Coherent (thick) and incoherent (thin lines) diffractive $J/\Psi$ production cross section as a function of $t$,
with (solid lines) and without (dashed lines) subnucleonic fluctuations. The band shows statistical uncertainty of the calculation.}
		\label{fig:t_spectra_5tev}
\end{figure}

The coherent and incoherent $J/\Psi$ production cross sections in ultraperipheral Pb+Pb collisions at $\sqrt{s_{NN}}=5.02$ TeV calculated using the Boosted Gaussian wave function are shown in Fig.~\ref{fig:t_spectra_5tev}. The cross sections at $y=0$ are first calculated without subnucleonic fluctuations, and then including both the geometric and $Q_s$ normalization fluctuations for the nucleons. Only the three first coherent peaks are shown, as 
calculating the average becomes numerically challenging at high~$|t|$~\cite{Toll:2012mb}. Further, it will be hard to measure the coherent cross section in this regime. 

We find that the coherent cross section is 
slightly reduced
when the subnucleon scale fluctuations are included. 
This change is caused by the modification of the impact parameter dependence of the dipole amplitude discussed in Sec.~\ref{sec:dipole}.

Below $|t|\lesssim 0.25\gev^2$, where one is sensitive only to fluctuations on length scales larger than the nucleon size, the incoherent cross section is approximately the same with and without nucleon structure fluctuations. On the other hand, at larger $|t|$ subnucleonic fluctuations clearly modify the slope of the incoherent cross section. The $|t|$ value $0.25\gev^2$, where the incoherent cross section becomes sensitive to subnucleonic fluctuations, corresponds to a distance scale $\sim 0.4 \fm$, which is of the order of the sizes of the gluonic hot spots in the nucleon (their root mean square \emph{radius} is $\sqrt{2 B_q} \approx 0.24$ fm). 
Preliminary ALICE data on exclusive $J/\Psi$ production~\cite{AliceUPCQM2017} show that the change in slope occurs around $\sqrt{-t}\approx p_T \sim 0.5 \gev$, which is in quantitative agreement with our results.

\begin{figure}[tb]
\centering
		\includegraphics[width=0.5\textwidth]{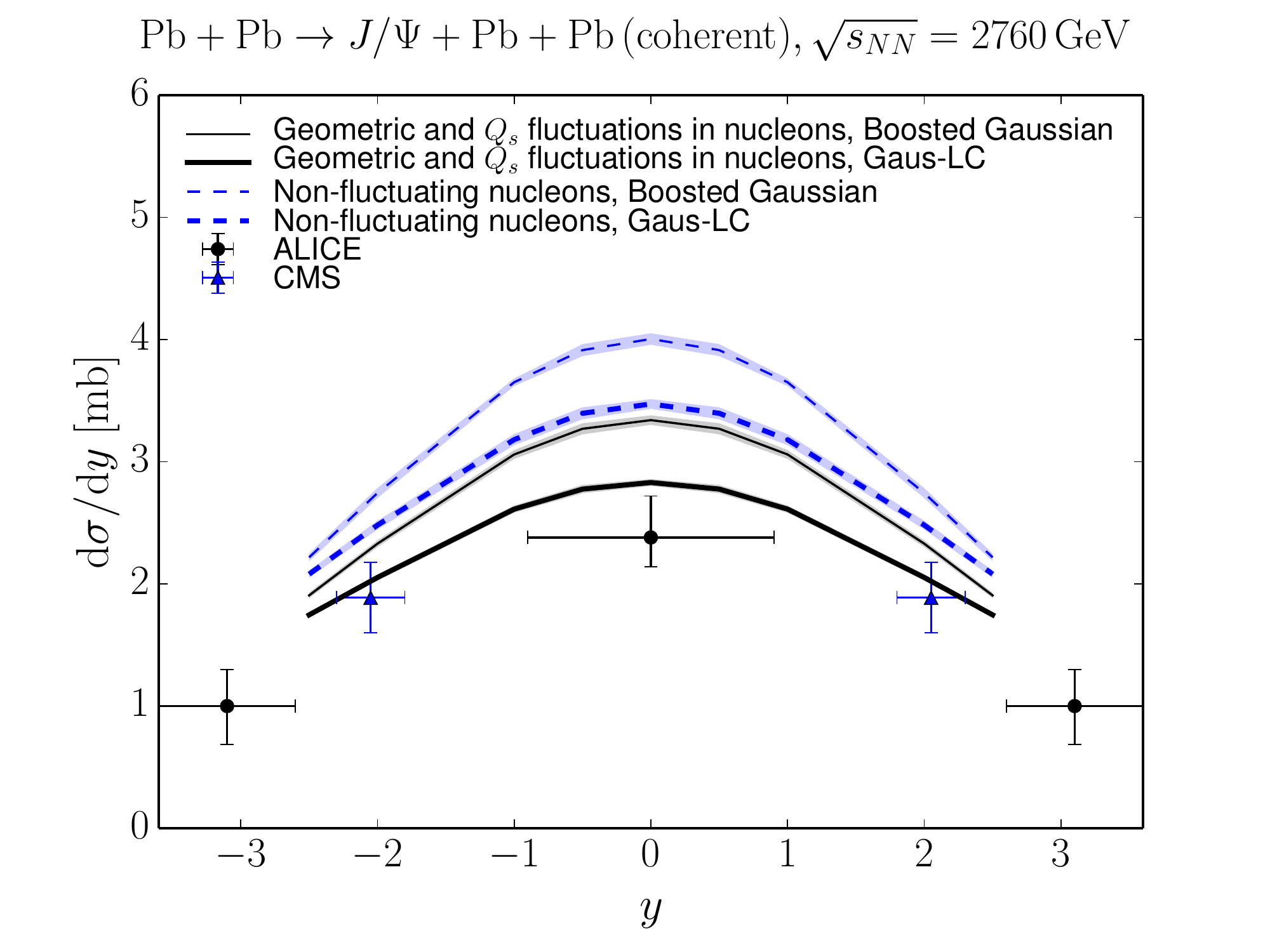} 
				\caption{Coherent $J/\Psi$ production cross section as a function of $J/\Psi$ rapidity with and without geometric fluctuations of the nucleon. Results are compared with the  ALICE~\cite{Abelev:2012ba,Abbas:2013oua} and CMS~\cite{CMS:2014ies}  data. 
				}
		\label{fig:coherent_2760}
\end{figure}

\begin{figure}[tb]
\centering
		\includegraphics[width=0.5\textwidth]{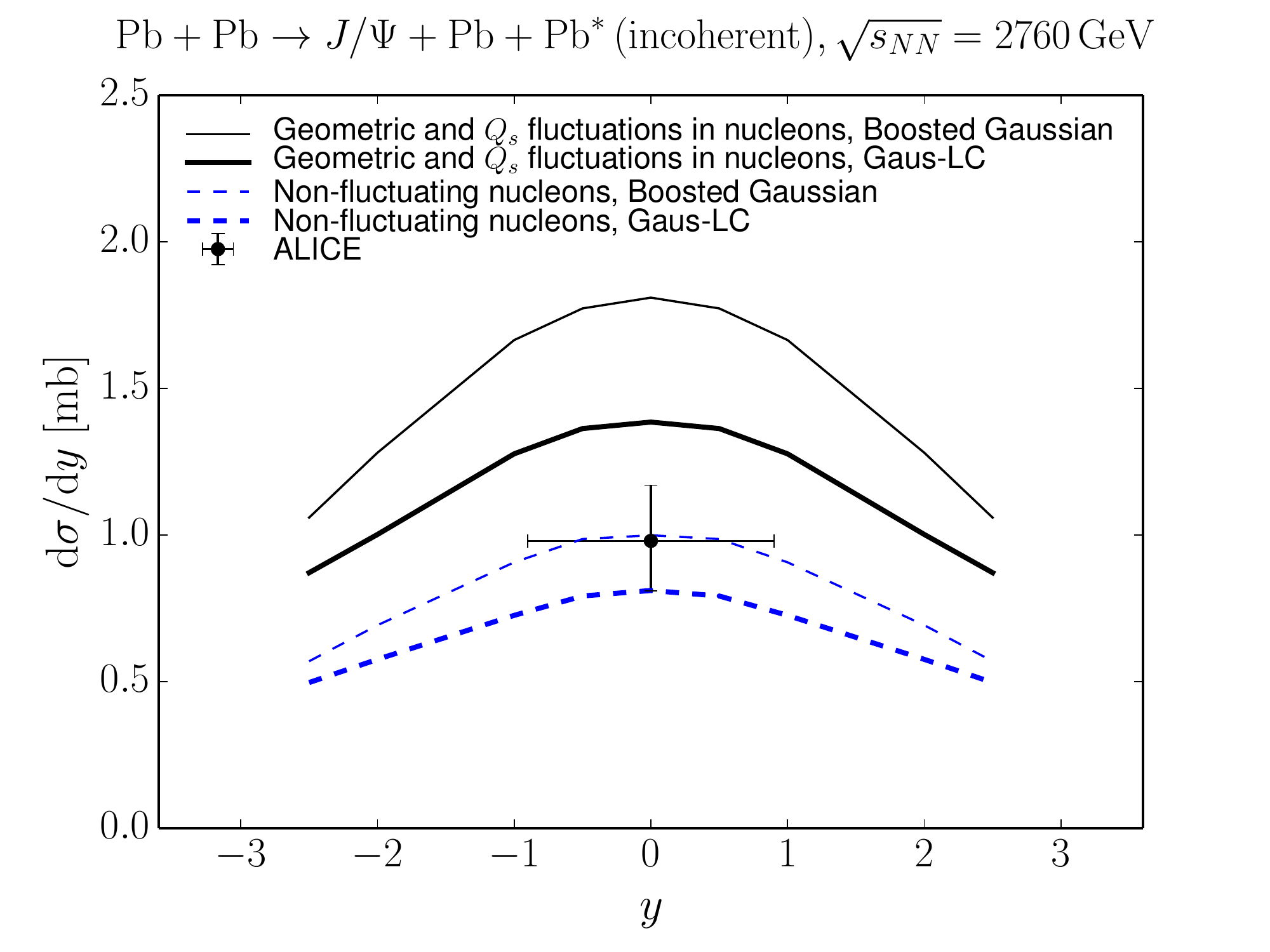} 
				\caption{Incoherent $J/\Psi$ production cross section as a function of $J/\Psi$ rapidity with and without geometric fluctuations of the nucleon. Experimental data from the ALICE collaboration~\cite{Abbas:2013oua}.}
		\label{fig:incoherent_2760}
\end{figure}

Next, we compare our results with the total ($t$ integrated) $J/\Psi$ production cross sections measured by ALICE~\cite{Abelev:2012ba,Abbas:2013oua} and CMS~\cite{CMS:2014ies} at $\sqrt{s_{NN}}=2.76$ TeV as a function of the $J/\Psi$ rapidity. The results for coherent and incoherent diffraction are shown in Figs.~\ref{fig:coherent_2760} and \ref{fig:incoherent_2760}, respectively. We show results obtained by using both the Boosted Gaussian and Gaus-LC wave functions for the $J/\Psi$. Similarly to previous literature~\cite{Lappi:2013am}, we find that the different wave functions mainly affect the overall normalization of the cross section.

In Fig.~\ref{fig:coherent_2760} we see that the coherent cross section is somewhat reduced when subnucleonic fluctuations are included. 
This change is comparable to the model uncertainties related to the $J/\Psi$ wave function.
In particular, we find that replacing the Boosted Gaussian by the Gaus-LC parametrization, both the coherent and incoherent (shown in Fig.\,\ref{fig:incoherent_2760}) cross sections are reduced by approximately 20\%.

Using the Boosted Gaussian wave function, without subnucleon scale fluctuations, the coherent cross section is significantly overestimated,  as in Ref.~\cite{Lappi:2013am}, and the incoherent cross section agrees with the data. Including geometric fluctuations increases the incoherent cross section almost by a factor of $2$, and the results are consistently  above the data for both processes. The rapidity dependence of the data is well reproduced. Results with subnucleon scale fluctuations obtained with the Gaus-LC wave function are close to the experimental data for the coherent cross section  and slightly higher for the incoherent cross section. At midrapidity, the ALICE coherent cross section datapoint is overestimated by $1.3\sigma$ and the incoherent cross section by $2\sigma$. We note that there is tension between the HERA $e+p$ data where the coherent cross section is underestimated by our model with subnucleonic fluctuations~\cite{Mantysaari:2016jaz}.

\begin{figure}[tb]
\centering
		\includegraphics[width=0.5\textwidth]{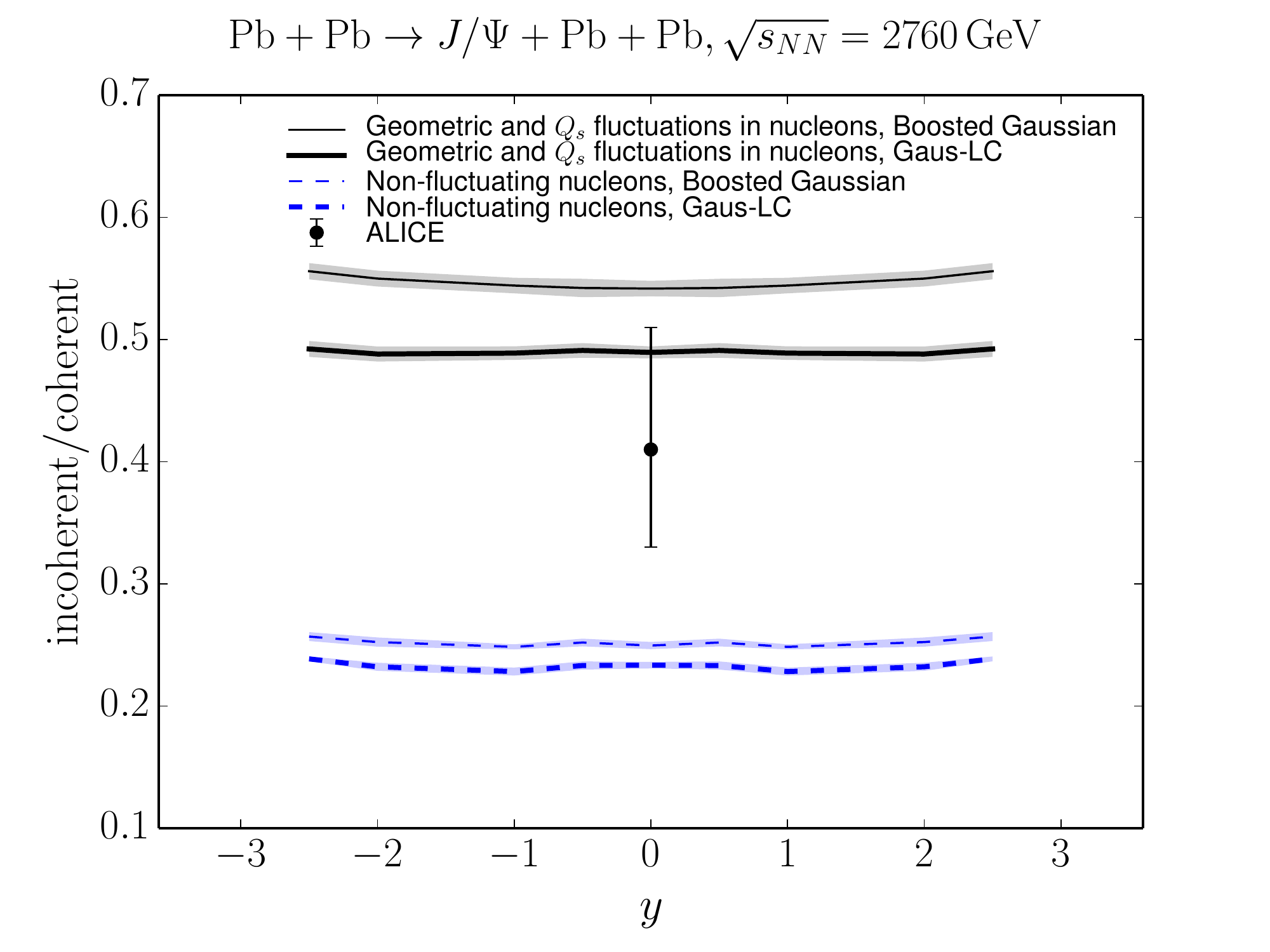} 
				\caption{Ratio of incoherent and coherent cross sections for the $J/\Psi$ production as a function of $J/\Psi$ rapidity with and without geometric fluctuations of the nucleon. Experimental data from the ALICE collaboration~\cite{Abbas:2013oua}.}
		\label{fig:ratio_2760}
\end{figure}

To reduce model uncertainties related to the vector meson wave function we study the ratio of incoherent to coherent cross section.
Our results are compared to the ALICE data~\cite{Abbas:2013oua} at $\sqrt{s_{NN}}=2.76$ TeV in Fig.~\ref{fig:ratio_2760}.
It can be seen that inclusion of the subnucleonic fluctuations brings this ratio to a level compatible with the experimental data. The ratio is found to be approximately independent of rapidity and we confirm that the dependence on the vector meson wave function is very weak.

\begin{figure}[tb]
\centering
		\includegraphics[width=0.5\textwidth]{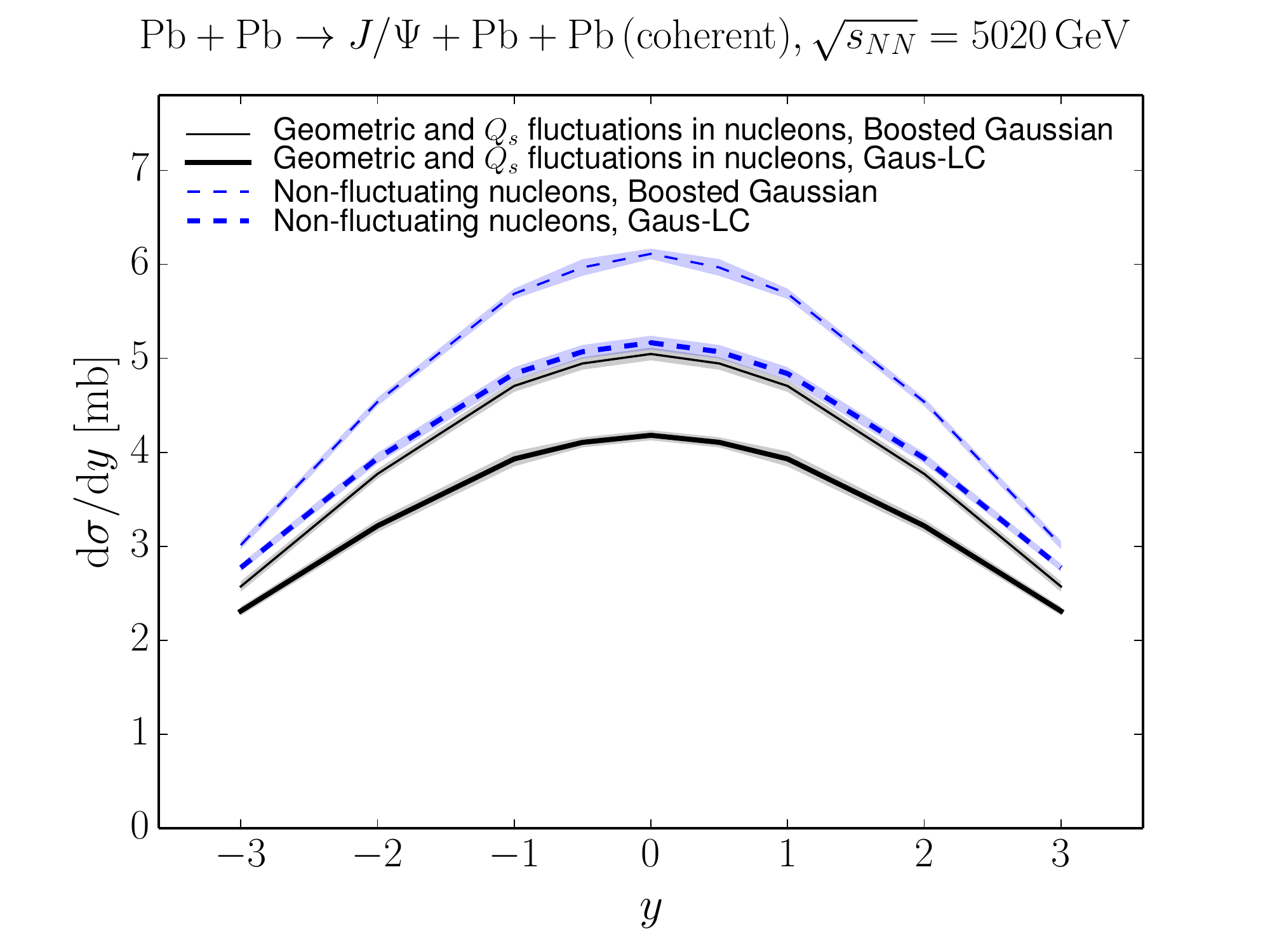} 
				\caption{Coherent $J/\Psi$ production cross section as a function of $J/\Psi$ rapidity with and without geometric fluctuations of the nucleon.}
		\label{fig:coherent_5020}
\end{figure}

\begin{figure}[tb]
\centering
		\includegraphics[width=0.5\textwidth]{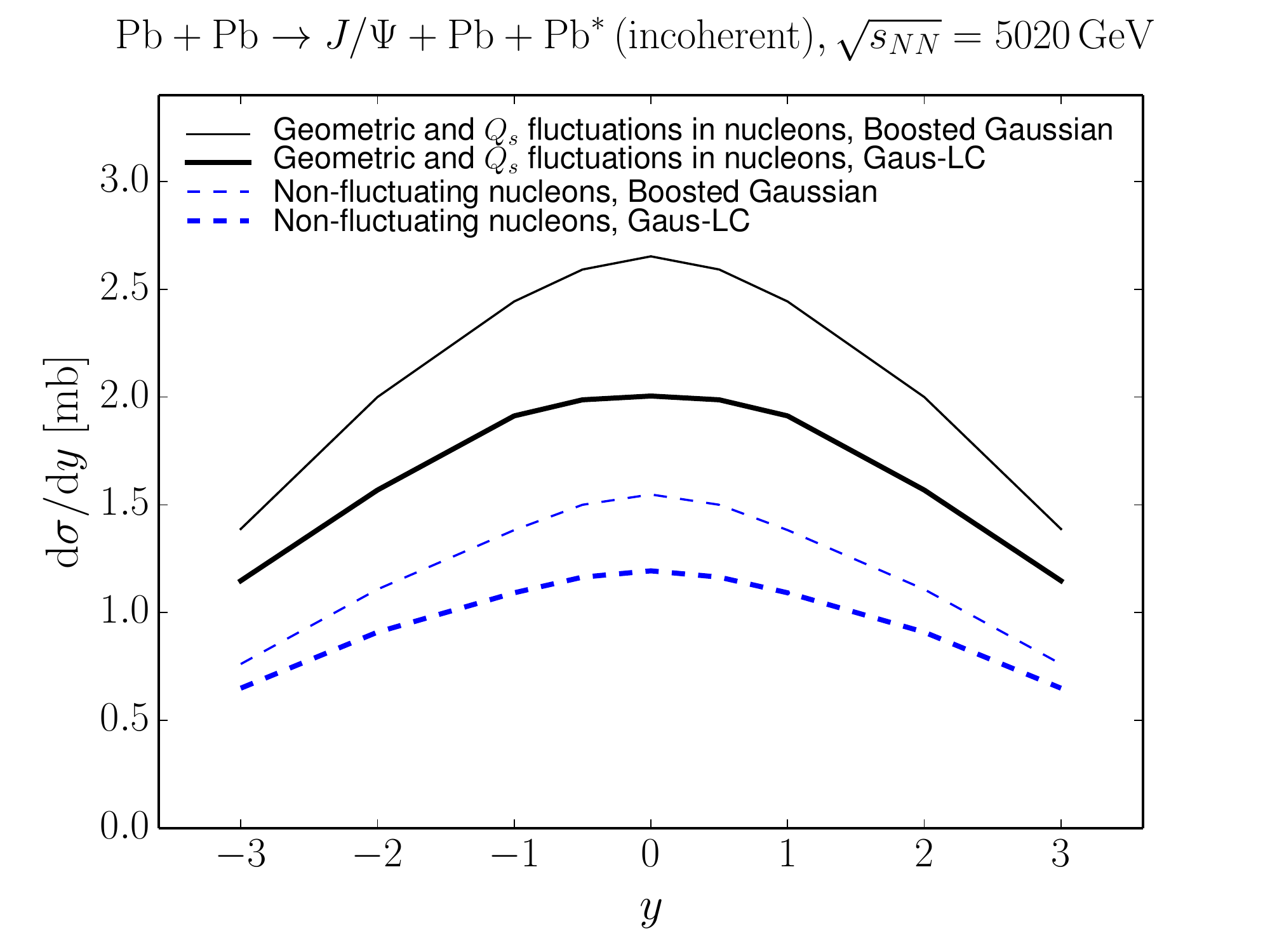} 
				\caption{Coherent $J/\Psi$ production cross section as a function of $J/\Psi$ rapidity with and without geometric fluctuations of the nucleon.}
		\label{fig:incoherent_5020}
\end{figure}

Predictions for the $J/\Psi$ rapidity distribution in ultraperipheral Pb+Pb collisions at  $\sqrt{s_{NN}}=5.02$ TeV are shown in Figs.\,\ref{fig:coherent_5020} and \ref{fig:incoherent_5020}. As for the lower energy, we find that the subnucleon scale geometric fluctuations have a large effect on the incoherent cross section. The ratio of the cross sections for the two processes for $\sqrt{s_{NN}}=5.02$ TeV decreases due to larger saturation effects on the incoherent cross section~\cite{Lappi:2010dd}. Numerically this effect is small in the LHC energy range, and we find that at $\sqrt{s_{NN}}=5.02$ TeV the ratio decreases by $0.5 \dots 3\%$.

\begin{figure}[tb]
\centering
		\includegraphics[width=0.5\textwidth]{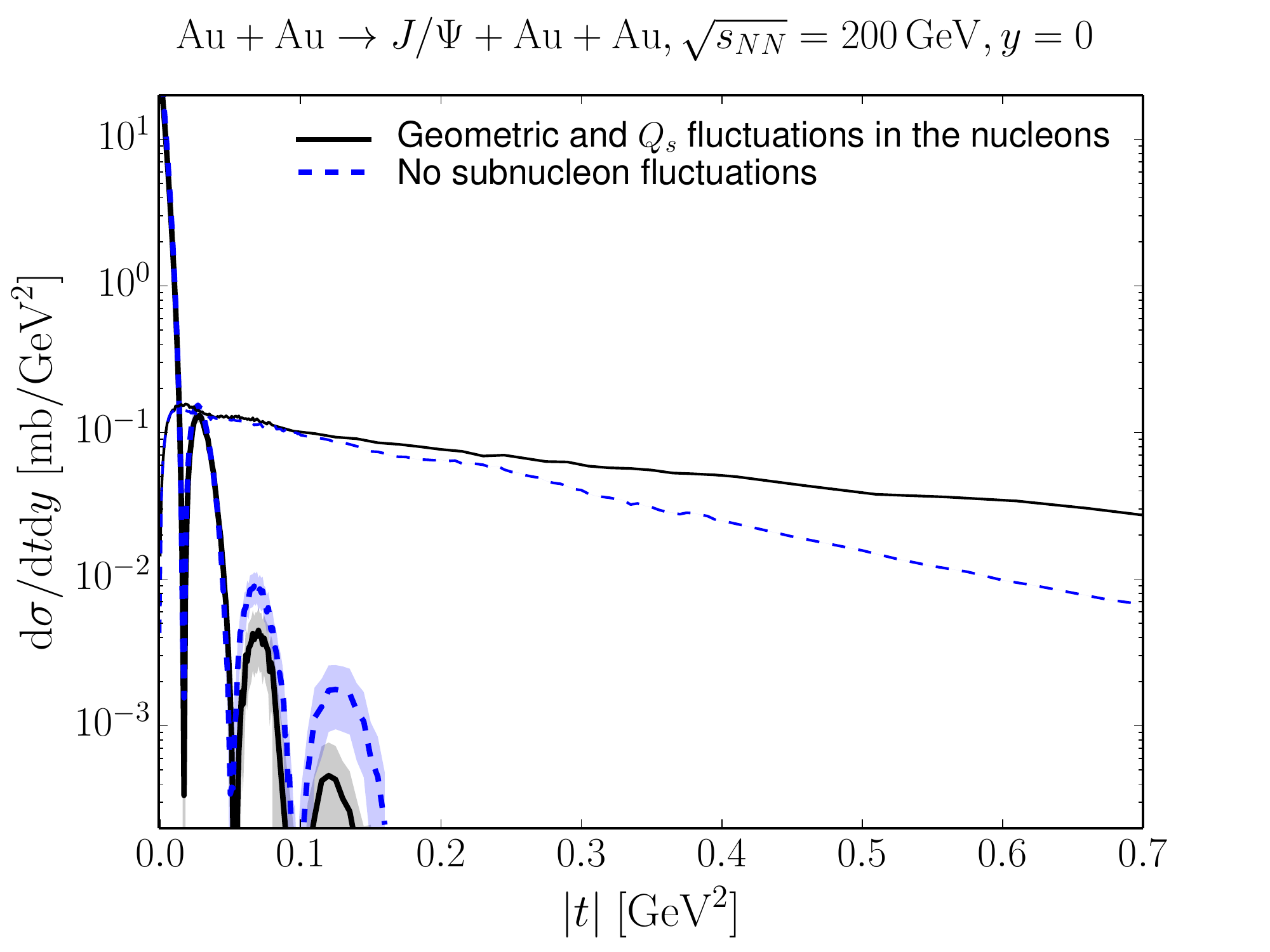} 
				\caption{Coherent (thick) and incoherent (thin lines) diffractive cross section for $J/\Psi$ production at midrapidity in ultraperipheral Au$+$Au collisions at $\sqrt{s_{NN}}=200$ GeV.}
				
		\label{fig:t_spectra_200gev}
\end{figure}

Finally, in Fig.\,\ref{fig:t_spectra_200gev} we show $|t|$-spectra for $J/\Psi$ production at midrapidity in ultraperipheral $\sqrt{s_{NN}}=200$ GeV Au+Au collisions at RHIC, corresponding to $W=25 \gev$. This corresponds to $\xpom=0.015$, which is at the edge of the validity of our model. Especially the skewedness and real part corrections together are almost $100\%$, which makes the absolute normalization unreliable (see Appendix \ref{appendix:corrections} and Fig.~\ref{fig:correction_w}). The spectra are calculated using the Boosted Gaussian wave function.  Integrating the cross sections over $t$, we get $106$ $\mu$b for the coherent and $62$ $\mu$b for the incoherent cross section with subnucleon fluctuations (with Gaus-LC wave function the cross sections are $100$ $\mu$b and $52$ $\mu$b). The corresponding cross section results without fluctuations are $121$ $\mu$b and $33$ $\mu$b ($118$ $\mu$b and $30$ $\mu$b with Gaus-LC wave function). The results for the coherent cross section are in agreement with the PHENIX~\cite{Afanasiev:2009hy} result  $76 \pm 33 \pm 11$ $\mu$b. 

The Bjorken-$x$ evolution in this work comes directly from the $Q_s$ evolution in the IPsat model. Thus, the amount of fluctuations and the size of the hot spots do not change as a function of $x$ or center-of-mass energy. In principle the characteristic length scales ($\sim Q_s^{-1}(x)$) depend on the energy and recent explicit calculations show that protons grow and fluctuations are reduced
 when Bjorken-$x$ is decreased~\cite{Schlichting:2014ipa,Cepila:2016uku}.
  If that is the case, we would expect the incoherent cross section to grow more slowly with energy than in our calculation.

\section{Conclusions}
We have shown that the subnucleon scale fluctuations, in particular geometric fluctuations of the nucleon shape, contribute significantly to the incoherent $J/\Psi$ production cross section at $|t|\gtrsim0.25\,{\rm GeV}^2$, measured in ultraperipheral heavy ion collisions at the LHC and RHIC. This is the first main result of this work: we expect to see the incoherent $t$ slope of the $J/\Psi$ production cross section to change at the value of $t$ corresponding to the distance scale of the subnucleonic fluctuations. Compared to the case where the only contribution to the fluctuations originates from fluctuating nucleon positions, the $|t|$ integrated incoherent cross section increases almost by a factor of $2$ when geometric fluctuations of the nucleon shape are included. 

When comparing to experimental data, the cleanest presented result is the ratio of the incoherent and coherent cross sections, which eliminates a large part of the model uncertainties. It increases by a factor of two when subnucleonic fluctuations are included. This is the second main result of this work, and explains previous tension between the employed dipole model and experimental data in Ref.~\cite{Lappi:2013am}, where one generally overestimated the coherent and underestimated the incoherent cross section.

In this work the energy evolution only affects the saturation scale of the nucleons. Our framework does not involve any possible smoothening of the nucleon or nucleus as one evolves towards small-$x$, discussed e.g. in Refs.~\cite{Schlichting:2014ipa,Cepila:2016uku}. As shown in \cite{Cepila:2016uku}, one could expect that the smoothening of the proton slows down the growth of the incoherent cross section with energy.
Including these effects in our calculation explicitly by solving JIMWLK evolution equations~\cite{JalilianMarian:1996xn,JalilianMarian:1997jx,JalilianMarian:1997gr,Iancu:2001md} as done in e.g. \cite{Schenke:2016ksl}, is left for future work.

\section*{Acknowledgments}
We thank T. Lappi, D. Takaki and  R. Venugopalan for discussions. This work was supported under DOE Contract No. DE-SC0012704. This research used resources of the National Energy Research Scientific Computing Center, which is supported by the Office of Science of the U.S. Department of Energy under Contract No. DE-AC02-05CH11231. BPS acknowledges a DOE Office of Science Early Career Award.

\appendix

\section{Phenomenological corrections}
\label{appendix:corrections}
The diffractive scattering amplitude \eq \eqref{eq:diff_amp} is 
obtained by assuming that the dipole amplitude is completely real, which leads to a fully imaginary diffractive scattering amplitude before the Fourier transform to  momentum space. 
The ratio between the real and the imaginary part of the amplitude, $\beta$, can be calculated as (see e.g.~\cite{Kowalski:2006hc})
\begin{equation}
\beta = \tan \frac{\pi \lambda}{2},
\end{equation}
where
\begin{equation}
\lambda = \frac{\der \ln \A^{\gamma^* A \to VA}}{\der \ln 1/\xpom}.
\end{equation}
In this work, we follow the prescription of Ref.~\cite{Lappi:2010dd} and calculate the effect of the corrections from photon-proton scattering, assuming that the correction has the same effect in photon-nucleus scattering. This correction is taken into account by multiplying the obtained cross sections by a factor $1+\beta^2$. 

The second phenomenological correction we include the  skewedness correction, which takes into account the fact that in the two-gluon exchange the gluons in the target are probed at different longitudinal momentum fractions $x_1 \ll x_2\approx \xpom$~\cite{Martin:1997wy,Shuvaev:1999ce, Martin:1999wb}.

In the IPsat model the collinear factorization gluon distribution $\xpom g(\xpom, \mu^2)$ is corrected to correspond to the off-diagonal (or skewed) distribution, which depends on both $x_1$ and $x_2$, by multiplying it by a skewedness factor $R_g$. Following the prescription of Ref.~\cite{Kowalski:2006hc} we get
 \begin{equation}
 \label{eq:skew}
 R_g = 2^{2\lambda_g+3} \frac{\Gamma(\lambda_g+5/2)}{\sqrt{\pi}\,\Gamma(\lambda_g+4)}
 \end{equation}
 with 
 \begin{equation}
 \lambda_g = \frac{\der \ln \xpom g(\xpom,\mu^2)}{\der \ln 1/\xpom}.
 \end{equation}
For photon-nucleus scattering the skewedness correction is approximated by calculating its effect on  the photon-proton scattering, and using the obtained correction factor.

\begin{figure}[tb]
\centering
		\includegraphics[width=0.5\textwidth]{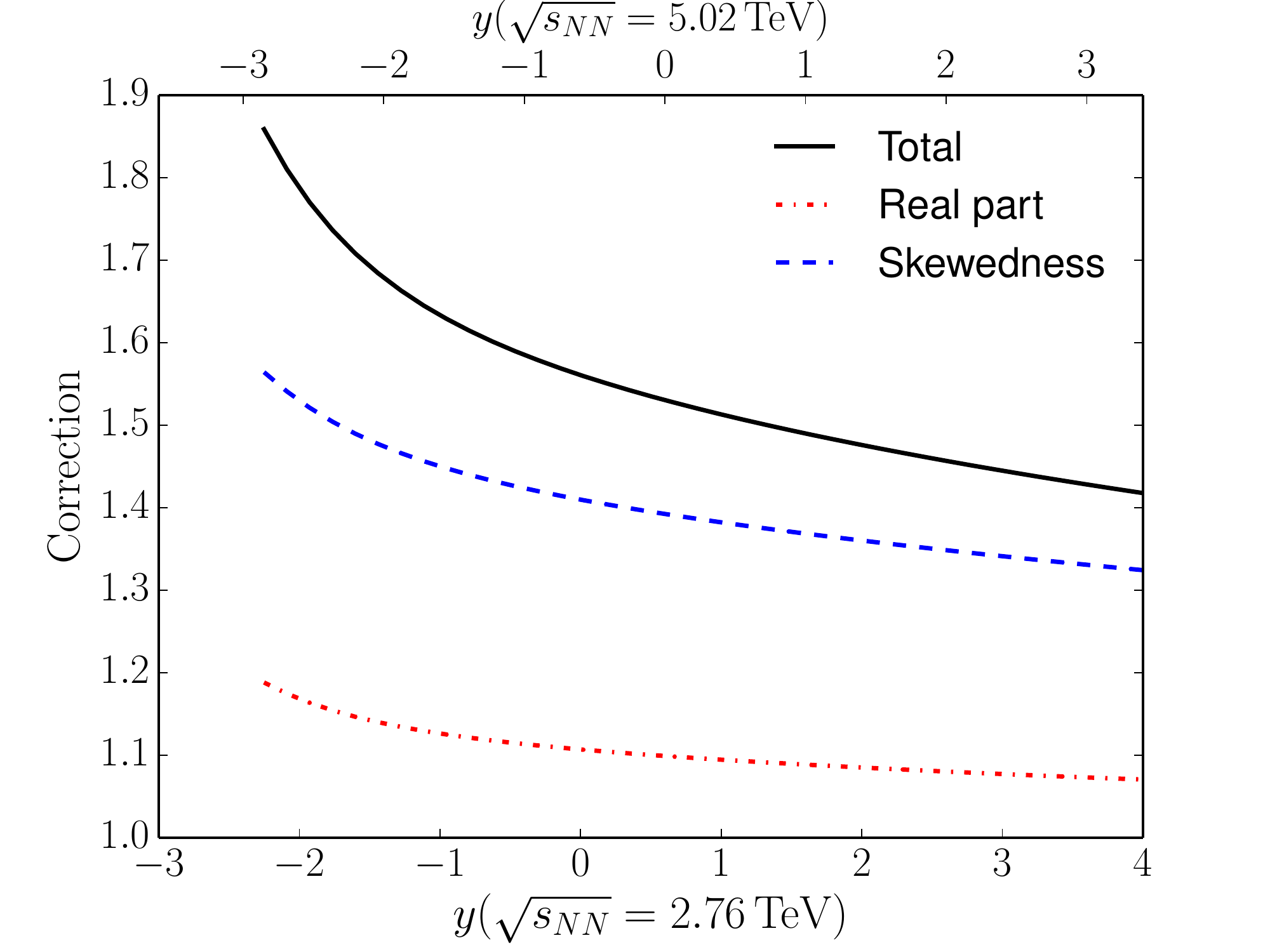} 
				\caption{Effect of the real part and skewedness correction as a function of rapidity $y$ in ultraperipheral collisions at two different center-of-mass energies. Negative rapidities refer to large-$x$ in the nucleus contribution, and positive values for the small-$x$ contribution.}
		\label{fig:correction_w}
\end{figure}

Especially the skewedness correction is numerically important and needed to describe the HERA diffractive measurements. The effect of real part and skewedness corrections on the total coherent diffractive cross section is shown in Fig.~\ref{fig:correction_w}. When corrections are calculated, no proton fluctuations are taken into account. The correction grows rapidly towards small rapidities (small $y$). Note that when $J/\Psi$ production is calculated at nonzero rapidity, there are always large-$x$ contributions (corresponding to negative $y$) and small-$x$ contributions (positive $y$) to the cross section, see Eq.~\eqref{eq:upc}.

\bibliographystyle{JHEP-2modlong.bst}
\bibliography{../../../refs}

\end{document}